# Monolithic tantalum pentoxide microrings with intrinsic Q factors exceeding 4×10$^6$

Xinzhi Zheng[a,b], Yixuan Yang[c,d], Renhong Gao[a,*], Lingling Qiao[c], Jintian Lin[c,d,*], and Ya Cheng[a,b,e,f,g,h,*]

[a] *State Key Laboratory of Precision Spectroscopy, East China Normal University, Shanghai 200062, China*

[b] *The Extreme Optoelectromechanics Laboratory (XXL), School of Physics, East China Normal University, Shanghai 200241, China*

[c] *State Key Laboratory of Ultra-intense laser Science and Technology, Shanghai Institute of Optics and Fine Mechanics, Chinese Academy of Sciences, Shanghai 201800, China*

[d] *Center of Materials Science and Optoelectronics Engineering, University of Chinese Academy of Sciences, Beijing 100049, China*

[e] *Shanghai Research Center for Quantum Sciences, Shanghai 201315, China*

[f] *Hefei National Laboratory, Hefei 230088, China*

[g] *Collaborative Innovation Center of Extreme Optics, Shanxi University, Taiyuan 030006, China*

[h] *Collaborative Innovation Center of Light Manipulations and Applications, Shandong Normal University, Jinan 250358, China*

**Abstract:** Tantalum pentoxide ($Ta_2O_5$), as a silicon-photonic-compatible material platform, has garnered significant attention for high-performance integrated photonics due to its exceptional properties: a broad transparency window spanning from 0.28 µm to 8 µm, a moderate refractive index of 2.05 at 1550 nm, and an impressive nonlinear refractive index of $7.2\times10^{-19}$ $m^2$/W. Despite these advantages, achieving low-loss fabrication of monolithic microrings on the $Ta_2O_5$ platform remains challenging due to its inherent hardness and brittleness, which often result in rough sidewalls and significant scattering losses. In this work, we successfully demonstrated monolithic $Ta_2O_5$ microring resonators with exceptionally high intrinsic and loaded quality (Q) factors. This was accomplished through the innovative application of photolithography-assisted chemo-mechanical etching (PLACE) technology. By optimizing the coupling region between the microring and the bus waveguide, as well as meticulously controlling surface roughness during fabrication, we achieved near-critical coupling in the resulting microrings. The devices exhibited loaded Q factors of $2.74\times10^6$ in the telecom band without employing expensive electron-beam lithography, showing an intrinsic Q factor as high as $4.47\times10^6$ and a low propagation loss of only 0.0732 dB/cm – representing the highest results reported for strongly confined $Ta_2O_5$-based microring resonators to date. This work paves the way for the development of advanced photonic devices on the $Ta_2O_5$ platform with low manufacturing cost, including low-threshold microlasers, highly sensitive sensors, broad bandwidth supercontinuum sources, and optical frequency combs.

[*]Corresponding author.

E-mail addresses: rhgao@phy.ecnu.edu.cn (R. Gao); jintianlin@siom.ac.cn (J. Lin); ya.cheng@siom.ac.cn (Y. Cheng).

## 1. Introduction

Tantalum pentoxide ($Ta_2O_5$) has recently emerged as a promising photonic integration platform due to its exceptional optoelectronic properties [1-6]. These properties include a large bandgap of 3.8 eV, a wide transparency window ranging from 280 nm to 8000 nm, a moderate refractive index of 2.05, a low thermo-optic coefficient ($5.75\times10^{-6}$ $K^{-1}$), and an impressively high nonlinear refractive index (~$7.2\times10^{-19}$ $m^2/W$), which is three times higher than that of $Si_3N_4$. These characteristics enable strong light confinement at sub-wavelength scales, dense integration, and broad operational bandwidth. Furthermore, $Ta_2O_5$ thin films can be deposited at room temperature with high throughput, offering potential for revolutionary advancements in various photonic applications such as on-chip supercontinuum generation, integrated frequency combs, optical true time delayed lines, highly sensitive biochemical sensing, and all-optical modulators [13].

Despite these remarkable advantages, $Ta_2O_5$ presents challenges due to its inherent hardness and brittleness, which complicate the fabrication of waveguide structures that simultaneously achieve low propagation loss and high confinement [14,15]. As a result, sophisticated electron-beam lithography (EBL) followed with reactive ion etching (RIE) is frequently required to suppress the sidewall roughness of the fabricated waveguide-based devices [11,14], which inevitably faces limitations in photonic integration scale without stitching error, restricted light-matter interaction length, low nonlinear conversion efficiency, and high optical power consumption.

In this work, monolithic integrated $Ta_2O_5$ microring resonators with ultra-high quality (Q) factors were successfully fabricated using the photolithography-assisted chemo-mechanical etching (PLACE) technique [16,17]. Near-critical coupling was achieved by optimizing the coupled region between the microring and the bus waveguide. The suppression of surface roughness to only 0.47 nm enabled monolithic microrings with an ultra-smooth surface. These structures exhibited a high loaded Q factor of $2.74\times10^6$ and an intrinsic Q factor of $4.47\times10^6$, corresponding to a low propagation loss of 0.0732 dB/cm. These Q factors and propagation loss represent the state-of-the-art performance on the $Ta_2O_5$ platform to date, without employing EBL. The successful realization of these strongly confined and low loss waveguides underscores the potential of $Ta_2O_5$ as a key material for next-generation integrated photonic applications.

## 2. Device design, fabrication, and experimental setup for performance characterization

To produce the high-quality $Ta_2O_5$ microrings, a $Ta_2O_5$ on insulator wafer was prepared by depositing a 600-nm-thick amorphous $Ta_2O_5$ thin film on a commercially available 4.7-μm-thick thermally grown silicon dioxide film on silicon wafer, through magnetron sputtering. Then, the femtosecond laser photolithography-assisted chemo-mechanical etching (PLACE) technique was leveraged to fabricate the high-Q monolithic microrings, as schematically illustrated in Figs. 1(a)-(d). First, a 600-nm-thick chromium (Cr) layer was deposited on the $Ta_2O_5$ film surface via magnetron sputtering, leveraging its high hardness as a hard mask for subsequent etching. Second, spatially selective ablation was carried out to pattern the Cr layer into microring structures side-coupled with the waveguides using the femtosecond pulsed laser. The microring was designed with a radius of 300 μm, a top width of 4.0 μm, and the coupled waveguide was designed with a width of 1.2 μm. Then, the Cr mask pattern was transferred to the $Ta_2O_5$ film via the chemo-mechanical polishing (CMP) process, by selectively removing the exposed $Ta_2O_5$ region and keeping the other region covered with the Cr hard mask intact. The residual Cr mask was then removed using a Cr etching solution. Finally, a secondary CMP was performed to further improve the surface smoothness of the $Ta_2O_5$ microring and reduce residual surface defects, leading to the formation of monolithic microrings with ultra-smooth surfaces and side walls.

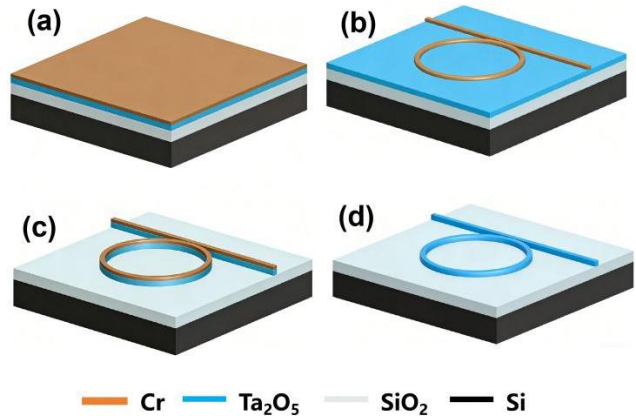

Fig. 1 Schematic of the fabrication flow of the microring resonators. (a) Coating a $Ta_2O_5$ layer and a chromium (Cr) layer on a silicon dioxide ($SiO_2$) film on silicon (Si) wafer successively. (b) Patterning the Cr film into micropattern which consists of microrings side-coupled with bus waveguides through femtosecond laser ablation to form hard mask. (c) Transferring the Cr mask into the $Ta_2O_5$ film by chemo-mechanical polishing (CMP). (d) Removing the Cr mask by chemical wet etching, and improving surface quality by secondary CMP.

The surface morphology and structure of the $Ta_2O_5$ microrings were characterized using scanning electron microscopy (SEM) and atomic force microscopy (AFM). Figure 2(a) shows the SEM image of the fabricated monolithically integrated microring side-coupled with a bus waveguide, exhibiting a smooth structure with distinct boundaries between the two coupled components. As presented in Figs. 2(b) and 2(c), the AFM characterization reveals an ultra-smooth microring surface with a root-mean-square (rms) roughness of only 0.46 nm. Such ultra-smooth surfaces are critical for minimizing waveguide scattering losses and enhancing the microcavity Q factor.

It is worth noting that the etched depth of the $Ta_2O_5$ structures in the side-coupling region is less than that in other waveguide regions. Considering that light can be coupled between the bus waveguide and the microring in the form of evanescent waves, the shallower etching depth in the side-coupling region relaxes the coupling distance requirement, provides additional design degree of freedom to boost efficient coupling, and increase the manufacturing tolerance. Here, the propagation of the optical field in the coupling region between the bus waveguide and the ring was simulated. As depicted in Fig. 2(d), even though the gap between the waveguide and the microring reaches as large as 8.7 μm, light can still be coupled from the waveguide to the microring, providing sufficient coupling efficiency.

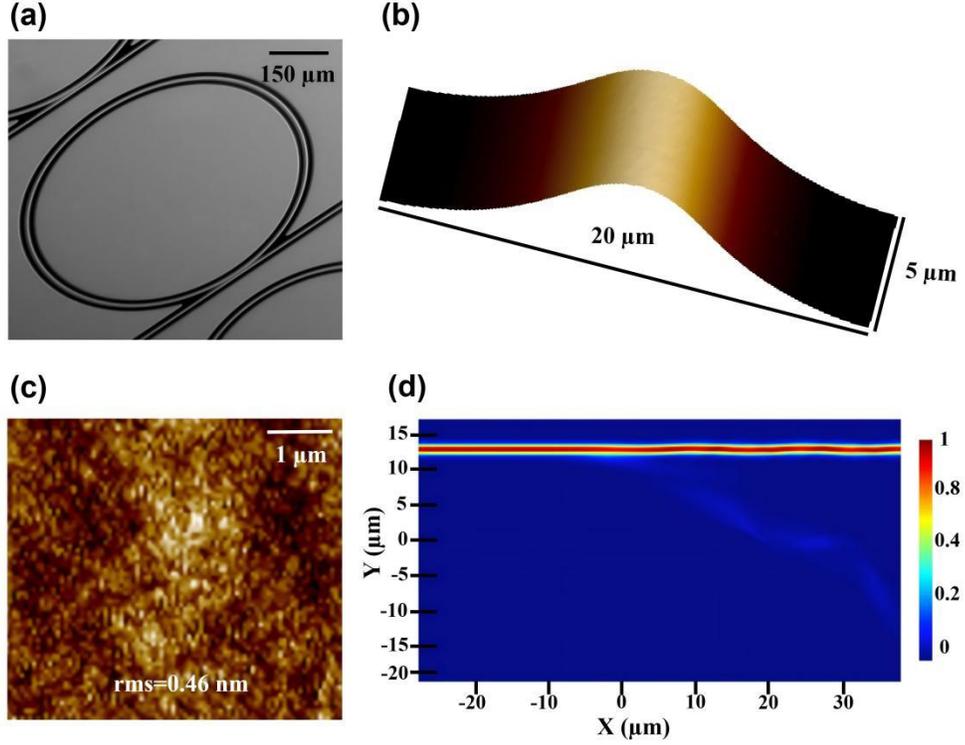

Fig. 2 The fabricated $Ta_2O_5$ microring. (a) SEM image of the microring. (b) AFM image of the microring waveguide. (c) AFM image of the microring surface. (d) Simulated field distribution of the microring coupling region.

To experimentally characterize the cavity resonances of the microring resonator, the experimental setup is schematically illustrated in Fig. 3. A tunable external-cavity diode laser (Model: TLB-6728, New Focus Inc.) was used as a pump light source with a linewidth < 200 kHz, scanned over the 1520-1570 nm wavelength range. An inline polarization controller (PC) was used to adjust the input light to be transverse-electrically (TE) polarized. The pump light was injected into the bus waveguide using a lensed fiber by end butt coupling, achieving a coupling efficiency of ~13%. Then the injected light was coupled into and out of the microring via evanescent wave. The on-chip laser power was set to 5 μW to avoid thermal and nonlinear optical effects. And the transmitted signal output from the bus waveguide was collected by another lensed fiber, and recorded by a photodetector (Model: 1611-FC, New Focus Inc.) and a digital oscilloscope (Model: MDO3104, Tektronix Inc.). To characterize the mode structure, the transmission spectrum spanning from 1520 to 1570 nm was measured by roughly scanning the laser wavelength through mechanically changing the external-cavity length of the tunable laser. And to measure the Q factor of each mode, a triangular-wave signal generator was used to drive the laser's high-resolution piezo-electric accessory for fine wavelength scanning around each specific resonant wavelength, exhibiting Lorentzian line-shape resonant dip for extracting the Q factor.

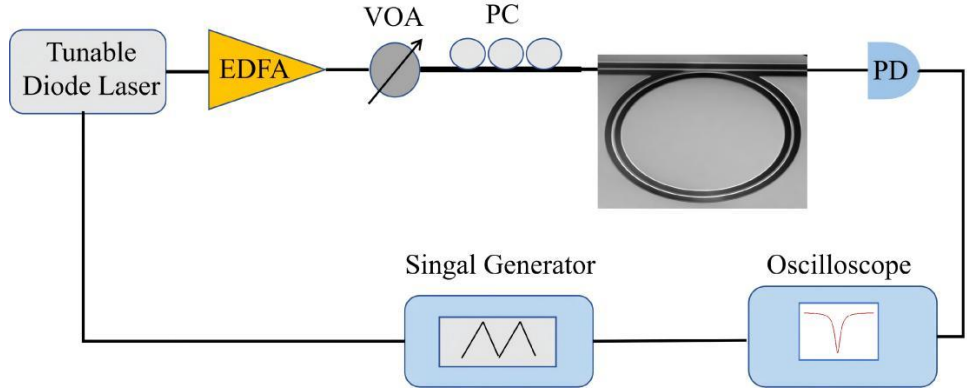

Fig. 3 Experimental setup for mode characterization. PD: photodetector. VOA: variable optical attenuator. PC: polarization controller.

## 3. Results

Figure 4(a) shows the transmission spectrum of the microring in the telecom band, exhibiting a uniform distribution of the fundamental TE polarized resonances combined with high-order spatial modes. Figure 4(b) plots the normalized transmission spectrum with a Lorentzian line-shape fit to one fundamental TE polarized mode, showing a loaded Q factor ($Q_L$) of $2.74 \times 10^6$, and near-critical coupling with a high transmission ($T_F$) of 95%. The intrinsic Q factor ($Q_i$) is calculated by the equation $Q_i = 2 \times Q_L/(1+T_F^{1/2})$, resulting in $Q_i = 4.47 \times 10^6$. The corresponding propagation loss of the waveguide is also calculated by the other equation $\alpha = 2\pi \cdot n_{eff} \cdot Q_i/\lambda$ (where $n_{eff}$=1.86, and $\lambda$ denote the effective refractive index and the resonant wavelength of the mode, respectively), demonstrating a low propagation loss of 0.0732 dB/cm. Table 1 summarizes the Q factors of the fabricated microring, along with a comparison with previously reported $Ta_2O_5$ microcavities. Our results represent the highest loaded and intrinsic Q factors for integrated $Ta_2O_5$ microcavities, which is attributed to the ultra-smooth sidewalls and nearly critical coupling. And the high Q factors in $Ta_2O_5$ microcavities significantly enhance the light field built up in the microring, enable a reduced pump threshold for soliton microcombs and improve sensitivity for sensing applications. Furthermore, this fabrication process can be extended to rare-earth-doped $Ta_2O_5$ films [18], facilitating lower pump thresholds for integrated microlasers and enhanced performance of on-chip amplifiers.

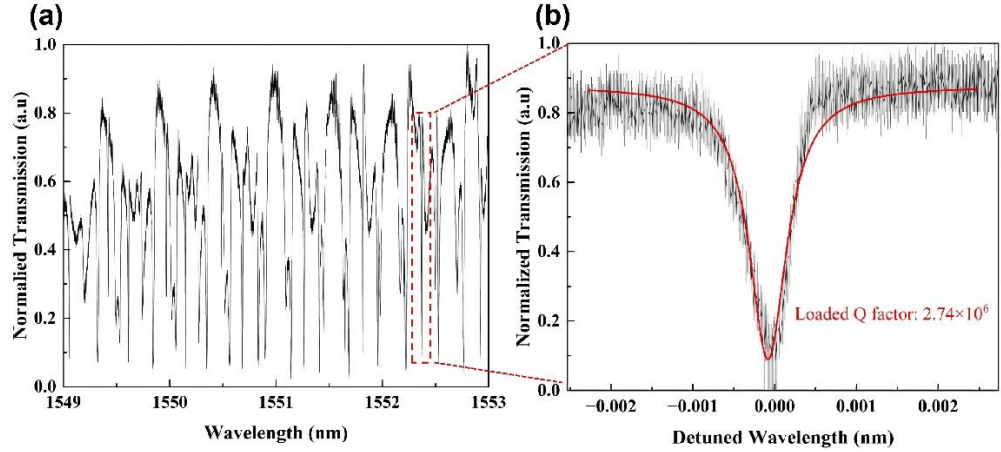

Fig. 4 (a) Transmission spectrum of the microring in the telecom C band. (b) The Lorentz fitting (red curve) reveals a loaded Q factor of $2.74 \times 10^6$, corresponding to an intrinsic Q factor of $4.47 \times 10^6$.

Table 1. Comparison of Q factors of on-chip $Ta_2O_5$ microcavities fabricated by different methods

| Microcavity Type | Loaded Q Factor | Intrinsic Q Factor | Fabrication Method | Reference |
| --- | --- | --- | --- | --- |
| Microring | $5.2 \times 10^5$ | $8.5 \times 10^5$ | EBL + RIE | [19] |
| Microring | $1.07 \times 10^6$ | $1.09 \times 10^6$ | EBL + RIE | [20] |
| Microring | $1.9 \times 10^6$ | $3.8 \times 10^6$ | EBL + RIE | [11] |
| Microdisk | $9.3 \times 10^5$ | $1.8 \times 10^6$ | EBL + RIE | [14] |
| Microring | $2.74 \times 10^6$ | $4.47 \times 10^6$ | PLACE | This Work |

## 4. Conclusion

In conclusion, we have demonstrated monolithically integrated $Ta_2O_5$ microring resonators with record-high Q factors via the PLACE technique. Such high Q factors are achieved by suppressing sidewall scattering losses via ultra-smooth surface fabrication with low manufacturing cost. This work addresses the key challenge of fabricating low-loss integrated $Ta_2O_5$ microresonators and opens up new opportunities for large-scale $Ta_2O_5$-based integrated photonic devices. The ultra-high Q $Ta_2O_5$ microrings can be used as core components for low-threshold microlasers [22-25], high-sensitivity sensors [26,27], optical amplifiers [28,29], frequency combs [30-32], and other high-performance photonic systems [33], and pave the way for fully photonic integrated circuits.